\documentclass[12pt]{article}
\usepackage{epsfig}
\usepackage{amsmath}
\usepackage{amssymb}
\usepackage{graphicx}
\usepackage{lscape}
\usepackage {subfigure}
\usepackage{hyperref, cite}
\def\be{\begin{equation}}
\def\ee{\end{equation}}
\def\ba{\begin{array}}
\def\ea{\end{array}}
\def\beqn{\begin{eqnarray}}
\def\eeqn{\end{eqnarray}}

\def\bt{\begin{tabular}}
\def\et{\end{tabular}}
\def\bc{\begin{center}}
\def\ec{\end{center}}
\usepackage{epstopdf}
\begin{document}
\title{ \textbf{CP odd weak basis invariants in minimal see-saw model and Leptogenesis}}
\author{Madan Singh$^{*} $\\
\it Department of Physics, M.N.S Government College Bhiwani, Bhiwani,\\
\it Haryana,127021, India.\\
\it $^{*}$singhmadan179@gmail.com
}
\maketitle
\begin{abstract}
In this paper, we derive the relationship between the weak basis invariants (WB) related to CP violation responsible for leptogenesis and CP violation relevant at low energy.  We  examine all the experimental viable cases of Frampton-Glashow and Yanagida (FGY) model, in order to construct  the  WB invariants in terms of left handed Majorana neutrino mass matrix elements, and thus finding the  necessary and sufficient condition for  CP conservation. Further for all the  viable FGY texture zeros, we have shown the explicit dependence of WB invariants on Dirac type and  Majorana type  CP violating phases. In the end, we discuss the implication of such interrelationships on leptogenesis.

\end{abstract}

\section{Introduction}
The  origin of CP violation  is one of the outstanding challenges in the fermion sector. In the Standard model (SM) \cite{S.M} CP violation is related to the mixing between the flavor and mass eigen states, also known as Cabibo-Kobayashi-Maskawa mechanism (CKM) \cite{cabibo} in the scenario of three families of quarks and non-degenerate masses, and is well established   in $K^{0}-\overline{K}^{0}$ system.
On the other hand, in the lepton sector, neutrinos are exactly massless Wely particle and lepton flavor mixing does not exist, implying that there is no CP violation in the lepton sector. However, several  neutrino oscillation experiments\cite{SNO,atm,react,accel}  provide us with very strong evidence regarding the non-zero neutrino masses as well as mixings.
  This, in consequence, provides  the first sign  to search for new physics and necessitates to  look  beyond the Standard model. In any extended model of SM, which incorporates neutrino masses and mixing, CP violation naturally appears  in the  leptonic sector. In the leptonic sector, CP violation  have profound implication in cosmology, playing a pivotal role in the generation of  matter-antimatter asymmetry of the universe via leptogenesis \cite{leptongensis1}. In this regard, seesaw mechanism \cite{seesa}, is  widely considered to be the most plausible candidate, which, not only, explain the smallness of neutrino masses in a natural way but also provides the origin of baryon asymmetry of the Universe. The seesaw mechanism, in fact, connects the small neutrino masses to very heavy right-handed neutrino
masses. In general it contains more physical parameters than can be measured at low energies. In an attempt to reduce the number of seesaw parameters, several theoretical ideas have been proposed  either by introducing the texture zeros in Yukawa coupling Dirac neutrino matrix or by reducing the right handed heavy Majorana neutrinos .
Among them, the most economical  is the imposition of two zeros in  Dirac neutrino mass matrix in the scheme of minimal seesaw model \cite{Ibrara, Casas}, popularly known as Frampton-Glashow- Yanagida (FGY) model \cite{FGY}. 
  However, the introduction of zeros in any specific model are not weak basis(WB) invariants, implying that a given set of texture zeros which exist in a certain WB may not be present or may appear in different entries in another WB, while leading to the same physics. This, in turn, brings forward a question of how to recognize the same texture zero model written in different bases where symmetry (or special texture zero) is not apparent. In such a scenario, CP odd weak basis invariants (WB) is considered to be an invaluable tool, and widely followed in the literature. The WB invariants were first used in \cite{Jarlskog} to study the CP violation in the quark sector. Similarly,  leptonic  WB invariants  were presented for  studying the  CP conditions  at low energy \cite{low1, low2, lh, invhl}.  To investigate the CP violation at high energies one requires  to establish a connection between the low energy physics and physics at high energies, for instance leptogenesis  \cite{lh, invhl}, \cite{pilaftsis,bridge,all}, and the  imposition of texture zeros  in the scenario of minimal seesaw model (MSM) may serve the purpose in this regard. This makes the study of CP-odd WB invariants relevant  for the model under consideration. In addition, it is crucial to examine the interrelationships between the CP-odd invariants which are required to vanish as a necessary and sufficient condition for CP conservation.

 The  present paper aims to study the implication of CP odd invariants for FGY ans\"{a}tze.   To this end, we first of all construct the CP-odd WB invariants relevant for leptogenesis (at high energies) in terms of left handed Majorana mass matrix elements at low energies for viable ans\"{a}tze, and then find the necessary and sufficient condition of CP conservation.  Further we  derive an analytical relations  showing an explicit dependence of the CP-odd invariants on Dirac/Majorana  CP violating phase. In the end we re-investigate the implications of such interrelationships on leptogenesis for each ans\"{a}tz.\\

\section{FGY ans\"{a}tze in minimal seesaw model}
In the present analysis, we take into account a most simple and economical see saw model \cite{Ibrara, Casas}, which  incorporates the two heavy right handed neutrinos $N_{1,2}$ having strong hierarchical pattern (i.e. $M_{2}>M_{1}$), and keep the Lagrangian of electroweak interactions invariant
under the $SU(2)_{L} \times U(1)_{Y}$ gauge transformation \cite{FGY}. After the spontaneous electroweak
symmetry breaking, this simple but interesting model leads to the following neutrino mass
term:
\begin{equation}\label{eq1}
-L_{mass}=(\overline{\nu_{e},\nu_{\mu},\nu_{\tau}})M_{D}\begin{pmatrix}
N_{1}\\
N _{2} 

\end{pmatrix}+\frac{1}{2}(\overline{N^{c}_{1},N^{c}_{2}})M\begin{pmatrix}
N_{1}\\
N _{2} 

\end{pmatrix} +\textit{h.c},
 \end{equation}
 where $N_{i}^{c}\equiv CN_{i}^{T} (i=1,2)$
 with $C$ being the charge-conjugation operator; and ($\nu_{e},\nu_{\mu},\nu_{\tau}$) denote the
left-handed neutrinos.  $M_{D}$ and $M$  denote a 3 $\times$ 2  Dirac neutrino mass  matrix,
and 2 $\times$ 2 symmetric Majorana neutrino mass matrix, respectively. The scale of
$M_{D}$ is characterized by the electroweak scale $v$ = 174 GeV. In contrast, the scale of $M$
can be much higher than $v$, because $N_{1}$ and $N_{2}$ are $SU(2)_{L}$ singlets and their corresponding
mass term is not subject to the scale of gauge symmetry breaking. Then one may obtain
the effective (light and left-handed) neutrino mass matrix $M_{\nu}$ via the well-known seesaw
mechanism \cite{seesa}

 \begin{equation}\label{eq2}
 M_{\nu}\approx M_{D}M^{-1}M_{D}^{T}.
 \end{equation}
  
 Without loss of generality, both heavy right-handed Majorana neutrino mass matrix $M$, and the charged lepton mass matrix $M_{l}$ are assumed 
to be diagonal, real and positive; i.e.,
\begin{equation} \label{eq3}
M=\left(
\begin{array}{ccc}
   M_{1}& 0& 0 \\
 0 & M_{2} & 0\\
  0& 0& 0 \\
  \end{array}
  \right) ,\; M_{l}=\left(
\begin{array}{ccc}
   m_{e}& 0& 0 \\
 0 & m_{\nu} & 0\\
  0& 0& m_{\tau} \\
  \end{array}
  \right),
  \end{equation}
where $M_{1,2}$ denotes the masses of two heavy Majorana neutrinos. The choice of this specific basis implies that one of the light (left-handed) Majorana neutrinos must be zero.
On the other hand,  $M_{D}$ is a complex 3 $\times$ 2 rectangular matrix, and can be given as
\begin{equation}\label{eq4}
 M_{D}=\left(
\begin{array}{cc}
   a_{1}& a_{2} \\
  b_{1} &b_{2} \\
  c_{1}& c_{2} \\
  \end{array}
  \right),
  \end{equation}
  where, $a_{1},a_{2},b_{1}, b_{2},c_{1}, c_{2}$ denote the complex entries.\\
  The minimal seesaw model itself has no restriction on the structure of $M_{D}$.  Frampton, Glashow and Yanagida \cite{FGY} first introduce the two zeros, with a aim to restrict the structure of $M_{D}$, whose origin comes from an underlying horizontal flavor symmetry.  Such ans\"{a}tze have been investigated  by many authors, while taking into account both  strongly hierarchical ($i.e.$  $M_{1}<< M_{2}$) \cite{Guo2002,Guo2006,32tex, Hargya} as well as nearly degenerate  ($i.e.$ $M_{1}\simeq M_{2}$) \cite{Zhou2015,schmitz2017} neutrino spectrum  of heavy right-handed Majorana neutrinos.   Among  the fifteen different  possibilities of Eq.(\ref{eq4}), only four are found to be compatible with neutrino oscillation data for inverted mas ordering, while same are ruled out for normal mass ordering \cite{Zhou2015}. The  four viable FGY ans\"{a}tze are given below: 
\begin{equation*} 
\rm{Type 1}:\quad M_{D}=\left(
\begin{array}{cc}
   a_{1}& 0 \\
  b_{1} &b_{2} \\
  0& c_{2} \\
  \end{array}
  \right),\quad
  \rm{Type 2}:\quad M_{D}=\left(
\begin{array}{cc}
   a_{1}& 0 \\
  0 &b_{2} \\
  c_{1}& c_{2} \\
  \end{array}
  \right);
  \end{equation*}
  \begin{equation}\label{eq5}
\rm{Type3}:\quad M_{D}=\left(
\begin{array}{cc}
   0& a_{2} \\
  b_{1} &0 \\
  c_{1}& c_{2} \\
  \end{array}
  \right),\quad
  \rm{Type 4}:\quad M_{D}=\left(
\begin{array}{cc}
   0& a_{2} \\
  b_{1} &b_{2} \\
  c_{1}&0  \\
  \end{array}
  \right).
  \end{equation}
  It is worthwhile to note here that in the MSM, the low-energy  phenomenological implications are driven by $M_{\nu}$, while cosmological baryon number asymmetry is associated with  $M_{D}$  via the leptogenesis mechanism.\\
  
  \section{Parameterization of lepton mass matrices in MSM}
  Before proceeding further, we briefly go through the  different parameterizations used for effective Majorana neutrino mass matrix ($M_{\nu})$ and Yukawa coupling Dirac neutrino mass matrix ($M_{D}$), respectively.  These may be useful for deriving the relationship between  CP odd invariants  related to CP violation at high energies and CP violation at low energies.
  As mentioned earlier,  the lightest neutrino in the MSM must be massless, therefore
we are then left with either $m_{1} = 0$ (normal mass ordering) or $m_{3} = 0 $(inverted
mass ordering). Since normal mass ordering is ruled out for all the FGY ans\"{a}tze, therefore we restrict our analysis for inverted mass ordering. In the basis of diagonal $M_{l}$, $M_{\nu}$ can be  parameterized  as follows 
\begin{equation}\label{eq6}
M_{\nu}\equiv\left(
\begin{array}{ccc}
   m_{ee}& m_{e \mu}& m_{e \tau} \\
  m_{e \mu} & m_{\mu \mu} & m_{\mu \tau}\\
  m_{e \tau}& m_{\mu \tau}& m_{\tau \tau} \\
  \end{array}
  \right)=V\left(
\begin{array}{ccc}
   m_{1}& 0& 0 \\
  0 & m_{2} & 0\\
  0& 0& 0 \\
  \end{array}
  \right)V^{T},
  \end{equation}
  
 where 
 \cite{116}
\begin{equation}\label{eq7}
V= \left(
\begin{array}{ccc}
 c_{12}c_{13}& s_{12}c_{13}& s_{13} \\
-c_{12}s_{23}s_{13}-s_{12}c_{23}e^{-i\delta} & -s_{12}s_{23}s_{13}+c_{12}c_{23}e^{-i\delta} & s_{23}c_{13}\\
 -c_{12}c_{23}s_{13}+s_{12}s_{23}e^{-i\delta}& -s_{12}c_{23}s_{13}-c_{12}s_{23}e^{-i\delta}& c_{23}c_{13} \\
\end{array}
\right)               
\left(
\begin{array}{ccc}
 1 & 0& 0 \\
0 & e^{i\sigma} & 0\\
0& 0& 1 \\
\end{array}
 \right).
\end{equation}
Here, $c_{ij} = \cos \theta_{ij}$, $s_{ij}= \sin \theta_{ij}$ for $i, j=1, 2, 3$, and $\delta$, $\sigma$ denote the Dirac and Majorana CP violating phase, respectively. 
From Eqs.(\ref{eq6}) and (\ref{eq7}), it is obvious that $M_{\nu}$ depends on seven low energy physical parameters: two neutrino masses ($m_{1},m_{2})$, three mixing angles ($ \theta_{12},\theta_{23}, \theta_{13})$,  two CP violating phases ($\delta,\sigma$), therefore, one can trivially derive each element of $M_{\nu}$ in terms of these parameters. The number of available parameters here, is lesser than that found  in $ M_{D}$, which reduces to nine after
eliminating the three trivial phases by rephasing the charged-lepton field in the chosen basis.
To account this difference, Casas-Ibarra-Ross \cite {Casas,Ibrara} introduce a orthogonal complex matrix R

\begin{equation}\label{eq8}
R=\left(
\begin{array}{cc}
   0& 0 \\
  cosz &-sinz \\
   sinz& cosz  \\
  \end{array}
  \right),
  \end{equation}\\
  for $m_{1}=0$, and for $m_{3}$ =0 is
  \begin{equation}\label{eq9}
R=\left(
\begin{array}{cc}
   cosz &-sinz \\
  sinz&cosz  \\
  0& 0 \\
  \end{array}
  \right).
  \end{equation}\\
  The complex parameter z encodes the two hidden parameters  $viz.$ a real parameter and one  phase, which are required  to match the total number of parameters at high energies and low energies in the MSM model.  
Using Eq.(\ref{eq2},\ref{eq8}, \ref{eq9}), one can now parameterize the $M_{D}$
in terms of V, $M_{\nu}, M$, and z as  
\begin{equation}\label{eq10}
M_{D}=iV\sqrt{m}R \sqrt{M}.
\end{equation} 
Using Eq.(\ref{eq10}),  for $ m_{3}=0$, one obtain

\begin{equation}\label{eq11}
a_{1}=i\sqrt{M_{1}}(V_{e1}\sqrt{m_{1}}c_{z} + V_{e2}\sqrt{m_{2}}s_{z}),  
\end{equation}
\begin{equation}\label{eq12}
a_{2}=i\sqrt{M_{1}}(V_{e2}\sqrt{m_{1}}s_{z} - V_{e3}\sqrt{m_{2}}c_{z}),
\end{equation} \\   
 where,  $V_{e1}, V_{e2}, V_{e3}$ denote the first row elements of neutrino mixing matrix  given in Eq.(\ref{eq7}). The remaining elements of $M_{D}$ can be expressed in the same manner  following the generic relations used in
    \cite {Casas,Ibrara}.

\section{Weak basis invariant(WB) for leptogenesis}

   In the seesaw mechanism,  lepton number asymmetry can be generated through the decays of the heavy Majorana neutrinos $M_{1}, M_{2}$. This is called leptogenesis mechanism \cite{leptongensis1, leptogensis2} and requires CP violation at high energies. Taking into account the  general seesaw mechanism, it is not possible to establish a connection between leptonic CP violation at low energies and CP violation at high energies. Such a relation can only be establish in the context of flavor theory.\
   Using the single flavor approximation for leptogenesis (i.e. in the case when wash out effects are not sensitive to the different flavors of the charged leptons into which the heavy neutrino decays), the  leptogenesis can be probed using the CP odd invariants \cite{pilaftsis}.
   
In the weak basis (WB), where  $M$ and $M_{l}$  are real and  diagonal, there are six physical phases in $M_{D}$, which  can be used to characterize the CP violation in the leptonic sector . This corresponds to  six     possible CP-odd WB invariants relevant for leptogenesis \cite{bridge}. For instance,
\begin{equation}\label{eq13}
I_{1}\equiv Im Tr[M_{D}^{\dagger}M_{D} (M^{\dagger}M) M^{*}(M_{D}^{\dagger}M_{D})^{*}M].
\end{equation}

The non-zero value of $I_{1}$ signals the CP violation in leptonic sector.  Since WB invariants are basis independent.  Therefore, in the chosen basis, one can express $I_{1}$ as
\begin{eqnarray}\label{eq14}
 &&\quad\qquad\quad\qquad I_{1}= M_{1}M_{2}(M_{2}^{2}-M_{1}^{2})Im [k_{12}^{2}]+ M_{1}M_{3}(M_{3}^{2}-M_{1}^{2})Im [k_{13}^{2}]
  \nonumber  \\ 
 &&~~~~~ 
 \quad\qquad \quad\qquad +M_{2}M_{3}(M_{1}^{3}-M_{2}^{2})Im [k_{23}^{2}],
 \end{eqnarray}
where, $k=M_{D}^{\dagger}M_{D}$ denotes the  $3\times3$ hermitian mass matrix. Clearly, $I_{1}=0$ implies CP conservation in leptonic sector.  This condition holds for either degenerate right-handed neutrino masses or diminishing imaginary part of $k_{ij}^{2}$ ($i \neq j$, i=1, 2, 3) or both. The interest in $I_{1}$ stems from the dependence on the term  Im ($k_{ij}^{2}$), which eventually determines the strength of leptogenesis. Hence  one can say that $I_{1}$ is  sensitive to the CP violating phases which appear in the leptogenesis.   

Following the similar WB as above,   CP-odd invariants $I_{2}$ and $I_{3}$ can  be expressed  as 

\begin{equation*}
I_{2}\equiv Im Tr[M_{D}^{\dagger}M_{D} (M^{\dagger}M)^{2} M^{*}(M_{D}^{\dagger}M_{D})^{*}M],
\end{equation*}
\begin{eqnarray}\label{eq15}
 &&\quad\qquad\quad\qquad 
=M_{1}M_{2}(M_{2}^{4}-M_{1}^{4})Im[k_{12}^{2}]+ M_{1}M_{3}(M_{3}^{4}-M_{1}^{4})Im[k_{13}^{2}]
\nonumber  \\ 
 &&~~~~~ 
 \quad\qquad \quad\qquad +M_{2}M_{3}(M_{3}^{4}-M_{2}^{4})Im[k_{23}^{2}].
\end{eqnarray}
and,
\begin{equation*}
I_{3}\equiv Im Tr[M_{D}^{\dagger}M_{D} (M^{\dagger}M)^{2} M^{*}(M_{D}^{\dagger}M_{D})^{*}M(M^{\dagger}M)],
\end{equation*}
\begin{eqnarray}\label{eq16}
 &&\quad\qquad\quad\qquad 
=M_{1}^{3}M_{2}^{3}(M_{2}^{2}-M_{1}^{2})Im[k_{12}^{2}]+ M_{1}^{3}M_{3}^{3}(M_{3}^{2}-M_{1}^{2})Im[k_{13}^{2}]
\nonumber  \\ 
 &&~~~~~ 
 \quad\qquad \quad\qquad 
+M_{2}^{3}M_{3}^{3}(M_{3}^{2}-M_{2}^{3})Im[k_{23}^{2}].
\end{eqnarray}

  In the MSM, one of the diagonal elements of $M$ ($i.e.$  $M_{3}=0$) is zero. Therefore, CP odd invariants in Eqs.(\ref{eq14},\ref{eq15},\ref{eq16}) are reduced to

\begin{equation} \label{eq17}
I_{1}=M_{1}M_{2}(M_{2}^{2}-M_{1}^{2})Im k_{12}^{2},
\end{equation}
\begin{equation} \label{eq18}
I_{2}=M_{1}M_{2}(M_{2}^{4}-M_{1}^{4})Imk_{12}^{2},
\end{equation}
\begin{equation}\label{eq19}
I_{3}=M_{1}^{3}M_{2}^{3}(M_{2}^{2}-M_{1}^{2})Imk_{12}^{2}.
\end{equation}
It must be noted that Eqs.(\ref{eq14}), (\ref{eq15}) and (\ref{eq16}) hold for the general case of  seesaw model, where all the three heavy Majorana neutrino masses ($M_{1}, M_{2}, M_{3}$) are real, diagonal and non-zero,  and  $M_{D}$ is $3\times3$ complex matrix. Hence $k$ turns out to be $3\times3$ hermitian matrix, while in minimal seesaw model, $M$ is a $2\times2$ real and diagonal matrix. This implies that $M_{D}$ is necessarily $3\times2$ complex matrix following the see-saw mechanism in Eq.(\ref{eq2}). Therefore, 
 $k$ is reduced to $2\times 2$  hermitian matrix  
 
\begin{equation*}
k_{11}=|a_{1}|^{2}+|b_{1}|^{2}+|c_{1}|^{2},\\
\end{equation*}
\begin{equation*}
k_{12}=a_{1}^{*} a_{2}+b_{1}^{*} b_{2}+c_{1}^{*} c_{2},\\
\end{equation*}
\begin{equation*}
k_{21}=a_{2}^{*} a_{1}+b_{2}^{*} b_{1}+c_{2}^{*} c_{1},\\
\end{equation*}
\begin{equation}\label{eq20}
k_{22}=|a_{2}|^{2}+|b_{2}|^{2}+|c_{2}|^{2}.\\
\end{equation}

The remaining three CP odd invariants $I_{4},I_{5}$ and $I_{6}$ can be written in a similar manner by simply substituting $M_{D}^{\dagger}M_{D}$ with $M_{D}^{\dagger}M_{l}M_{l}^{\dagger}M_{D}$  
\begin{equation*}
I_{4}\equiv Im Tr[M_{D}^{\dagger}M_{l}M_{l}^{\dagger}M_{D} (M^{\dagger}M) M^{*}(M_{D}^{\dagger}M_{l}M_{l}^{\dagger}m_{D})^{*}M],
\end{equation*}
\begin{eqnarray}\label{eq21}
 &&\quad\qquad\quad\qquad =  M_{1}M_{2}(M_{2}^{2}-M_{1}^{2})Im [K_{12}^{2}]+ M_{1}M_{3}(M_{3}^{2}-M_{1}^{2})Im [K_{13}^{2}]
  \nonumber  \\ 
 &&~~~~~ 
 \quad\qquad \quad\qquad +M_{2}M_{3}(M_{3}^{3}-M_{2}^{2})Im [K_{23}^{2}].
 \end{eqnarray}
\begin{equation*}
I_{5}\equiv Im Tr[M_{D}^{\dagger}M_{l}M_{l}^{\dagger}M_{D} (M^{\dagger}M)^{2} M^{*}(M_{D}^{\dagger}M_{l}M_{l}^{\dagger}m_{D})^{*}M],
\end{equation*}
\begin{eqnarray}\label{eq22}
 &&\quad\qquad\quad\qquad 
=M_{1}M_{2}(M_{2}^{4}-M_{1}^{4})Im[K_{12}^{2}]+ M_{1}M_{3}(M_{3}^{4}-M_{1}^{4})Im[K_{13}^{2}]
\nonumber  \\ 
 &&~~~~~ 
 \quad\qquad \quad\qquad +M_{2}M_{3}(M_{3}^{4}-M_{2}^{4})Im[K_{23}^{2}].
\end{eqnarray}
\begin{equation*}
I_{6}\equiv Im Tr[M_{D}^{\dagger}M_{l}M_{l}^{\dagger}M_{D} (M^{\dagger}M)^{2} M^{*}(M_{D}^{\dagger}M_{l}M_{l}^{\dagger}M_{D})^{*}M(M^{\dagger}M],
\end{equation*}
\begin{eqnarray}\label{eq23}
 &&\quad\qquad\quad\qquad 
=M_{1}^{3}M_{2}^{3}(M_{2}^{2}-M_{1}^{2})Im[K_{12}^{2}]+ M_{1}^{3}M_{3}^{3}(M_{3}^{2}-M_{1}^{2})Im[K_{13}^{2}]
\nonumber  \\ 
 &&~~~~~ 
 \quad\qquad \quad\qquad 
+M_{2}^{3}M_{3}^{3}(M_{3}^{2}-M_{2}^{3})Im[K_{23}^{2}].
\end{eqnarray}
Eqs. (\ref{eq21}), (\ref{eq22}) and (\ref{eq23})  can be deduced in MSM model as
 
 \begin{equation}\label{eq24}
I_{4}=M_{1}M_{2}(M_{2}^{2}-M_{1}^{2})Im K_{12}^{2},
\end{equation}
\begin{equation}\label{eq25}
I_{5}=M_{1}M_{2}(M_{2}^{4}-M_{1}^{4})Im K_{12}^{2},
\end{equation}
\begin{equation}\label{eq26}
I_{6}=M_{1}^{3}M_{2}^{3}(M_{2}^{2}-M_{1}^{2})Im K_{12}^{2}.
\end{equation}

where, K is $2\times2$ hermitian matrix and its elements are given below:

\begin{equation*}
K_{11}=m_{e}^{2}|a_{1}|^{2}+m_{\mu}^{2}|b_{1}|^{2}+m_{\tau}^{2}|c_{1}|^{2}\\,
\end{equation*}
\begin{equation*}
K_{12}=m_{e}^{2}a_{1}^{*} a_{2}+m_{\mu}^{2}b_{1}^{*} b_{2}+m_{\tau}^{2}c_{1}^{*} c_{2}\\,
\end{equation*}
\begin{equation*}
K_{21}=m_{e}^{2}a_{2}^{*} a_{1}+m_{\mu}^{2}b_{2}^{*} b_{1}+m_{\tau}^{2}c_{2}^{*} c_{1}\\,
\end{equation*}
\begin{equation} \label{eq27}
K_{22}=m_{e}^{2}|a_{2}|^{2}+m_{\mu}^{2}|b_{2}|^{2}+m_{\tau}^{2}|c_{2}|^{2}.\\
\end{equation}
where, $m_{e}, m_{\mu}$ and $m_{\tau}$ denote the electron, muon and tau neutrinos, respectively.\\
In the following section, we shall discuss the implications of six  CP odd invariants for FGY ans\"{a}tze.

  \section{Implication of CP-odd WB invariants for FGY ans\"{a}tze}
  
     As discussed in section 3,  $M_{\nu}$ consists of seven physical parameters.
Since $M_{\nu}$ is related to $M_{D}$ and $M$ through the seesaw relation, given in Eq. (\ref{eq2}), therefore,  the
parameters of $M_{\nu}$ are depend on $M_{D}$ and $M$. In principle,
the light Majorana neutrino masses, flavor mixing angles and CP-violating phases
can  all be  calculated at low energies. Hence it is possible to reconstruct $M_{D}$ by the means of two heavy Majorana neutrino masses ($M_{1}, M_{2}$) and the complex elements of $M_{\nu}$.   In the following discussion,  we  derive the CP-odd WB invariants  in terms of  $M_{1}, M_{2}$, and the complex elements of $M_{\nu}$  for  FGY ans\"{a}tze.  
  \subsection{Type 1}
Using the seesaw mechanism in Eq.(\ref{eq2}) and Eq. (\ref{eq5}),  one can write the expression for $M_{\nu}$ for type 1 as 
  \begin{equation}\label{eq28}
M_{\nu}=\left(
\begin{array}{ccc}
  \frac{a_{1}^{2}}{M_{1}}& \frac{a_{1}b_{1}}{M_{1}}& 0 \\
  &\frac{b_{1}^{2}}{M_{1}}+\frac{b_{2}^{2}}{M_{2}} &\frac{b_{2}c_{2}}{M_{1}} \\
  && \frac{c_{2}^{2}}{M_{2}} \\
  \end{array}
  \right),
  \end{equation}
  in terms of Dirac neutrino matrix elements $a_{1}, b_{1} , b_{2}, c_{2}$.
 On comparing Eqs.(\ref{eq6}) and (\ref{eq28}), one can trivially find $a_{1}, b_{1} , b_{2}, c_{2}$ in terms of the  elements of  Eq.(\ref{eq28}),  $a_{1}^{2}=M_{1}m_{ee}$, $b_{1}^{2}=M_{1} \frac{(m_{e\mu})^{2}}{m_{ee}}$, $c_{2}^{2}=M_{2}m_{\tau \tau}$, $b_{2}^{2}=\frac{M_{2}(m_{\mu \tau})^{2}}{m_{\tau\tau}}$.
Since $I_{i}$ ($i=1,2,3$) is directly proportional to Im $k_{12}^{2}$. Therefore it is sufficient to evaluate  $I_{1}$ for each FGY ans\"{a}tz. \\
Using Eq.(\ref{eq17}),  one can write $I_{1}$, for type 1, 
\begin{equation} \label{eq29}
I_{1}=M_{1}^{2}M_{2}^{2}(M_{2}^{2}-M_{1}^{2})Im\bigg[\frac{(m_{e\mu}^{*})^{2}m_{\mu\tau}^{2}}{m_{ee}^{*}m_{\tau\tau}}\bigg],
\end{equation}
 where, $k_{12}^{2}=(b_{1}^{*}b_{2})^{2}$. The CP violation depends on the phase i.e. $arg\bigg(\frac{(m_{e\mu}^{*})^{2}m_{\mu\tau}^{2}}{m_{ee}^{*}m_{\tau\tau}}\bigg)$. The vanishing of this phase implies
 CP conservation,  and leads to following phase relation
\begin{equation}\label{eq30}
arg(m_{ee})+2arg(m_{\mu\tau})=arg(m_{\tau\tau})+2arg(m_{e\mu}).
\end{equation}
 From the above equation, one can say that  CP violation is brought about by the mismatch among the phases of elements $m_{e\mu}^{2}$, $m_{\mu\tau}^{2}$, $m_{ee}$ and $m_{\tau\tau}$, while phase of the elements $m_{e\tau}$ or $m_{\mu\mu}$ does not have any contribution for CP violation and can be rephased away. 
 \subsection{Type 2}
  For type 2,   using  Eqs.(\ref{eq2}) and (\ref{eq5}), we get  
  \begin{equation}\label{eq31}
M_{\nu}=\left(
\begin{array}{ccc}
  \frac{a_{1}^{2}}{M_{1}}& 0& \frac{a_{1}c_{1}}{M_{1}} \\
  &\frac{b_{2}^{2}}{M_{2}} &\frac{b_{2}c_{2}}{M_{2}} \\
  &&\frac{c_{1}^{2}}{M_{1}}+\frac{c_{2}^{2}}{M_{2}}  \\
  \end{array}
  \right).
  \end{equation}
     Again, using Eq.(\ref{eq31}), one can easily find the following relations,  $a_{1}^{2}=M_{1}m_{ee}$, $c_{1}^{2}=M_{1} \frac{(m_{e\tau})^{2}}{m_{ee}}$, $c_{2}^{2}=M_{2} \frac{m_{\mu \tau}^{2}}{m_{\mu\mu}}$, $b_{2}^{2}=M_{2}m_{\mu \mu}$.
  
Using these relations and Eq.(\ref{eq17}),  one can derive $I_{1}$ for type 2, 
\begin{equation} \label{eq32}
I_{1}=M_{1}^{2}M_{2}^{2}(M_{2}^{2}-M_{1}^{2})Im\bigg[\frac{(m_{e\tau}^{*})^{2}m_{\mu\tau}^{2}}{m_{ee}^{*}m_{\mu \mu}}\bigg],
\end{equation}\\
where, $k_{12}^{2}=(c_{1}^{*}c_{2})^2$, and  CP violation  explicitly depends on physical phase $i.e.$  $arg\bigg(\frac{(m_{e\tau}^{*})^{2}m_{\mu\tau}^{2}}{m_{ee}^{*}m_{\mu\mu}}\bigg)$.\\
The necessary and sufficient condition for CP conservation for type 2 is given as
\begin{equation} \label{eq33}
arg(m_{ee})+2arg(m_{\mu\tau})=arg(m_{\mu\mu})+2arg(m_{e\tau}).
\end{equation} 
The type 1 and type 2 are phenomenologically related to each other via  $\mu-\tau$ exchange symmetry.
 \subsection{Type 3}
  Like type 2, type 3 also leads to $m_{e\mu}=0$, and  using Eqs.(\ref{eq2}) and (\ref{eq5}) ,  one gets 
  \begin{equation}\label{eq34}
M_{\nu}=\left(
\begin{array}{ccc}
  \frac{a_{2}^{2}}{M_{2}}& 0& \frac{a_{2}c_{2}}{M_{2}} \\
  &\frac{b_{1}^{2}}{M_{1}} &\frac{b_{1}c_{1}}{M_{2}} \\
  &&\frac{c_{1}^{2}}{M_{1}}+\frac{a_{2}c_{2}}{M_{2}}  \\
  \end{array}
  \right).
  \end{equation}
   Using Eqs.(\ref{eq6}) and (\ref{eq34}), we obtain the following mathematical relations for the elements of $M_{D}$:  $b_{1}^{2}=M_{1}m_{\mu\mu}$, $c_{1}^{2}=M_{1} \frac{(m_{\mu\tau})^{2}}{m_{\mu\mu}}$, $c_{2}^{2}=M_{2} \frac{m_{e \tau}^{2}}{m_{ee}}$, $a_{2}^{2}=M_{2}m_{ee}$.
   
 Using these relations, one can find 
\begin{equation}\label{eq35}
I_{1}=M_{1}^{2}M_{2}^{2}(M_{2}^{2}-M_{1}^{2})Im\bigg[\frac{(m_{\mu\tau}^{*})^{2}m_{e\tau}^{2}}{m_{\mu \mu}^{*}m_{ee}}\bigg],
\end{equation}\\
where, $k_{12}^{2}=(c_{1}^{*}c_{2})^2$, and CP violation for type 3 depends on the physical phase $i.e.$ $arg \bigg[\frac{(m_{\mu\tau}^{*})^{2}m_{e\tau}^{2}}{m_{\mu \mu}^{*}m_{ee}}\bigg]$, and its vanishing value leads to the following phase relation
\begin{equation}\label{eq36}
arg(m_{\mu\mu})+2arg(m_{e\tau})=arg(m_{ee})+2arg(m_{\mu\tau}).
\end{equation}
The results obtained here are just the complex conjugate of the results obtained in case of type 2.
   
\subsection{Type 4}
   For type4, using Eq.(\ref{eq2}) and (\ref{eq6}), one can write 
  \begin{equation} \label{eq37}
M_{\nu}=\left(
\begin{array}{ccc}
  \frac{a_{2}^{2}}{M_{2}}& \frac{a_{2}b_{2}}{M_{2}}&0  \\
  &\frac{b_{1}^{2}}{M_{1}}+\frac{b_{2}^{2}}{M_{2}} &\frac{b_{1}c_{1}}{M_{2}} \\
  &&\frac{c_{1}^{2}}{M_{1}}  \\
  \end{array}
  \right).
  \end{equation}
  Similar to type 1, type 4 also leads to $m_{e\tau}$=0.  
   Using Eq.(\ref{eq37}), we arrive at the following relations,  $b_{1}^{2}=M_{1} \frac{(m_{\mu\tau})^{2}}{m_{\tau\tau}}$, $c_{1}^{2}=M_{1}m_{\tau\tau}$, $b_{2}^{2}=M_{2} \frac{m_{e \mu}^{2}}{m_{ee}}$, $a_{2}^{2}=M_{2}m_{ee}$,
 and consequently, we obtain 
\begin{equation}\label{eq38}
I_{1}=M_{1}^{2}M_{2}^{2}(M_{2}^{2}-M_{1}^{2})Im\bigg[\frac{(m_{\mu\tau}^{*})^{2}m_{e\mu}^{2}}{m_{\tau \tau}^{*}m_{ee}}\bigg],
\end{equation}
where, $k_{12}^{2}=(b_{1}^{*}b_{2})^2$,
and, for CP conservation, we require
\begin{equation}\label{eq39}
arg(m_{\tau\tau})+2arg(m_{e\mu})=arg(m_{ee})+2arg(m_{\mu\tau}).
\end{equation}
  Like in type 1 and type 2, we  find that  type 3 and type 4 are also related  via $\mu-\tau$ exchange symmetry. In addition, the results obtained in type 3  are simply a complex conjugate to that in type 1, and the same is true for type 2 and type 4 texture zeros.
  Similarly, one can derive the relations for  $I_{2}$ and  $I_{3}$ in terms of Majorana mass matrix elements using Eqs. (\ref{eq18},\ref{eq19}). The  CP violating phase remain similar to $I_{1}$, while the coefficients dependence in terms of  heavy right handed neutrinos are different as shown in Eqs.(\ref{eq17}, \ref{eq18},\ref{eq19}).\\
 On the other hand, the remaining CP-odd invariants ($I_{4}, I_{5}, I_{6}$) depend on Im $K_{12}^{2}$.
 For illustration, we shall only evaluate $I_{4}$ for type 1.\\ 
 Using Eq.(\ref{eq24}), and elements of  $M_{D}$ provided in subsection 5.1, it is trivial to find the expression for $I_{4}$ 
\begin{equation}\label{eq40}
I_{4}=m_{\mu}^{2} M_{1}^{2}M_{2}^{2}(M_{2}^{2}-M_{1}^{2})Im\bigg[\frac{(m_{e\mu}^{*})^{2}m_{\mu\tau}^{2}}{m_{ee}^{*}m_{\tau\tau}}\bigg],
\end{equation}
where, $K_{12}^{2}=m_{\mu}^{2}(b_{1}^{*}b_{2})^{2}$.
The above relation is  similar to Eq. (\ref{eq29}) except that $I_{4}$ depends on additional charged lepton parameter $m_{\mu}$. The  CP invariance condition obtained here is  similar to $I_{1}$ for type 1. For the sake of completion, we have tabulated  all the CP-odd invariants for all the viable FGY ans\"{a}tze alongwith the  necessary and sufficient CP invariance condition in Table[\ref{tab1}]. The conditions on phases can be visualized as fine tuning required to have CP conservation at high energies.\\  
 From the above discussion, it is trivial to find that  $M_{D}$ with  three or more zeros leads to CP invariance in the leptonic sector. In addition, it is found that that  all the  CP-odd invariants strongly depend on the effective neutrino mass term $|m_{ee}|$ $i. e.$ $I_{i} \propto \frac{1}{|m_{ee}|^{2}}$, where $i=1, 2, 3, 4, 5, 6$.  If $|m_{ee}|$ =0 , $I_{i}$ simply blows up. Therefore the measurement of $|m_{ee}|$ in neutrinoless double beta decay experiments could have serious implications on these WB invariants. 

\section{CP-odd WB invariants and low energy CP violating phases}
In this section, we discuss how the CP odd invariants  depend on CP violating phases ($\delta, \sigma$) in an explicit manner. 
Using Eqs.(\ref{eq11}) and (\ref{eq12}), we get the following relations
 \begin{equation}\label{eq41}
 tanz=R^{1/2}t_{12}e^{i\sigma},
 \end{equation},
 \begin{equation} \label{eq42}
 cotz=-R^{1/2}t_{12}e^{i\sigma},
 \end{equation}\\
 for $a_{1}=0$ and $a_{2}=0$, respectively. The symbols $R=\frac{m_{2}}{m_{1}}$, and $t_{12}=\frac{c_{12}}{s_{12}}$.\\

 Using Eq.(\ref{eq41}), it is trivial to find the imaginary part as  
 \begin{equation}\label{eq43}
\rm{ Im[c_{z}^{2}]}=\frac{Rt_{12}^{2}sin2\sigma}{(1+Rt_{12}^{2}cos2\sigma)^{2}+(Rt_{12}^{2}sin2\sigma)^{2}},
 \end{equation}
 where $c_{z}=cos(z)$.\\
 For Type 1 and Type 4, we obtain,  $m_{e\tau}=0$ as evident from Eqs. (\ref{eq28}) and (\ref{eq37}).
 Using this constraint, we arrive at the following relation between $\delta$ and $\sigma$
 \begin{equation} \label{eq44}
 s_{13}sin\delta=-\frac{t_{12}t_{23}}{c_{12}^{2}}\times\frac{R sin2\sigma}{(1+Rt_{12}^{2}cos2\sigma)^{2}+(Rt_{12}^{2}sin2\sigma)^{2}}.
 \end{equation}
   Following the same procedure as in Eq.(\ref{eq43}),  we get, using Eq.(\ref{eq42}),
 \begin{equation}\label{eq45}
  \rm{Im[s_{z}^{2}]}=-\frac{Rt_{12}^{2}sin2\sigma}{(1+Rt_{12}^{2}cos2\sigma)^{2}+(Rt_{12}^{2}sin2\sigma)^{2}}.
 \end{equation}
  
For Type 2 and Type 3, we obtain, $m_{e\mu}=0$. Using this condition, one can easily obtain the  relation
 \begin{equation}\label{eq46}
 s_{13}sin\delta=\frac{t_{12}}{t_{23}c_{12}^{2}}\times\frac{R sin2\sigma}{(1+Rt_{12}^{2}cos2\sigma)^{2}+(Rt_{12}^{2}sin2\sigma)^{2}}.
 \end{equation}
 
 On comparing  Eqs. (\ref{eq43}) and (\ref{eq44}),
  we get  
 \begin{equation}\label{eq47}
 Im[c_{z}^{2}]=+t_{12}t_{23}^{-1}c_{12}^{2}s_{13}sin\delta,
 \end{equation}
 and, 
 \begin{equation}\label{eq48}
 Im[s_{z}^{2}]=+t_{12}t_{23}^{-1}c_{12}^{2}s_{13}sin\delta,
 \end{equation}
 for type1 and type4, respectively.\\
 Similarly, on comparing Eqs.(\ref{eq45}) and (\ref{eq46}), we get 
 \begin{equation}\label{eq49}
 Im[c_{z}^{2}]=-t_{12}t_{23}c_{12}^{2}s_{13}sin\delta,
 \end{equation}
 and, 
 \begin{equation}\label{eq50}
 Im[s_{z}^{2}]=-t_{12}t_{23}c_{12}^{2}s_{13}sin\delta,
 \end{equation}
 for type 2 and type 3, respectively.
 
 To evaluate the CP-odd invariants in terms of Dirac CP violating phase ($\delta$), we need to calculate the 
  $2\times 2$ hermitian matrix $k$. Using Eq.(\ref{eq10}), we have, $k=\sqrt{M}R^{\dagger}mR\sqrt{M}$.
  On solving it, we obtain\cite{Hargya},
  
  \begin{equation}\label{eq51}
 k
  =\left(
\begin{array}{cc}
   M_{1}(m_{1}|c_{z}|^{2}+m_{2}|s_{z}|^{2})& \sqrt{M_{1}M_{2}}(-m_{1}c_{z}^{*}s_{z}+m_{2}s_{z}^{*}c_{z}) \\
  \sqrt{M_{1}M_{2}}(-m_{1}c_{z}s_{z}^{*}+m_{2}s_{z}c_{z}^{*}) &M_{2}(m_{1}|s_{z}|^{2}+m_{2}|c_{z}|^{2}) \\
  \end{array}
  \right).
  \end{equation}\\
After squaring the above matrix, one can extract the term 
\begin{equation}\label{eq52}
k_{12}^{2}=\sqrt{M_{1}M_{2}}[(M_{1}m_{1}+M_{2}m_{2})|c_{z}|^{2}+(M_{1}m_{2}+M_{2}m_{1})|s_{z}|^{2}](-m_{1}c_{z}^{*}s_{z}+m_{2}s_{z}^{*}c_{z}).
\end{equation}
  \\
Since we know that imaginary part of pure imaginary number is again imaginary. Therefore,  one can write $I_{1}$ using Eq.(\ref{eq52}) 
 
  \begin{equation}\label{eq53}
  I_{1}\simeq CIm(s_{z}^{*}c_{z}), 
  \end{equation}
 where, $C\simeq 2M_{1}^{5/2}M_{2}^{5/2}(M_{2}^{2}-M_{1}^{2})(M_{1}+M_{2})m^{2}$, is the coefficient of $I_{1}$, and we have used the approximation,   $m_{1}\simeq m_{2}\simeq m$.
  Clearly, $I_{1}$ depends on the neutrino mass $m$ and heavy right-handed Majorana neutrino masses $M_{1}$ and $M_{2}$. The CP violation depends on the phase of complex term $s_{z}^{*}c_{z}$. 
  On evaluating further using Eq.(\ref{eq41}), we obtain 
   \begin{equation}\label{eq54}
    I_{1} \simeq Ct_{12}R^{1/2}|c_{z}|^{2}sin\sigma.
   \end{equation} 
 The relation holds for type1 and type2. \\   
For Type3 and Type4,  using Eq.(\ref{eq42}), $I_{1}$ is 
given as  
 \begin{equation}\label{eq55}
  I_{1}\simeq \frac{C}{t_{12}R^{1/2}}|c_{z}|^{2}sin\sigma.
\end{equation}
Similarly, we can easily derive the expressions for $I_{2}$, $I_{3}$, $I_{4}$, $I_{5}$ and $I_{6}$ in terms of sin$\sigma$ using Eqs. (\ref{eq18},\ref{eq19}, \ref{eq24}, \ref{eq25}, \ref{eq26})for each ansatz. 
From Eqs.(\ref{eq54}) and (\ref{eq55}), we conclude that  sin$\sigma$=0 leads to CP conservation in leptonic sector. Taking into account the analytical relation between $\delta$ and $\sigma$ in Eqs.(\ref{eq44}),(\ref{eq45}), one find that CP conservation holds for $ \rm{\delta, \sigma=  \pm n\pi}$, where n is a integer.

\section{Relationship between the thermal leptogenesis and left handed Majorana neutrino mass matrix}

In the thermal leptogenesis in the MSM, seesaw mechanism  with only two right handed neutrinos succeeds in reproducing the observed  baryon asymmetry of universe for a nearly degenerate heavy neutrino mass spectrum.  
In \cite{Dror},   seesaw mechanism with thermal leptogenesis is also tested in the context of gravitational waves.  D. Croon et.al\cite{croon} have studied how the observed baryon asymmetry is realized after high scale reheating into the lightest sterile neutrino in the framework of MSM.

In this choosen framework, the decays of two heavy right-handed Majorana neutrinos, $N_{i}\rightarrow l+H$ and
$N_{i}\rightarrow \overline{l}+H^{*}$ (for $i = 1, 2$), are both lepton-number-violating and CP-violating \cite{leptogensis2}. The CP asymmetry $\epsilon_{i}$ originates from the interference between the tree-level and one-loop
decay amplitudes. If $N_{1}$ and $N_{2}$ have a hierarchical mass spectrum ($M_{1} << M_{2}$), the
interactions involving $N_{1}$ can be in thermal equilibrium when $N_{2}$ decays. The asymmetry term $\epsilon_{2}$
is erased before $N_{1}$
 decays. The CP-violating asymmetry $\epsilon_{1}$, which is produced by
the out-of-equilibrium decay of $N_{1}$, in the choosen basis where $M_{l}$ and $M$ are both diagonal,  can be given as

\begin{equation*}
\epsilon_{1} \equiv \frac{\Gamma(N_{1}\rightarrow l+H)-\Gamma(N_{1}\rightarrow \overline{l}+H^{*})}{\Gamma(N_{1}\rightarrow l+H)+\Gamma(N_{1}\rightarrow \overline{l}+H^{*})},
\end{equation*}

\begin{equation}\label{eq56}
\approx -\frac{3}{16\pi v^{2}}\times \frac{M_{1}}{M_{2}} \times \frac{Im(k^{\dagger}k)_{12}^{2}}{(k^{\dagger}k)_{11}},
\end{equation}
In this section, we discuss the
implications of FGY ansatz  on leptogenesis.
 To this end, we find the relationship between $\epsilon_{1}$  and $M_{\nu}$ for each ansatz.\\
Using Eq.(\ref{eq56}) and  $a_{1}, b_{1} , b_{2}, c_{2}$ in subsection 5.1, one can easily arrive at, 
\begin{equation}\label{eq57}
\epsilon_{1}=-\frac{3}{16\pi v^{2}}\times \frac{M_{1}}{|m_{ee}||m_{\tau\tau}|^{2}(|m_{ee}|^{2}+|m_{e\mu}|^{2})} \times Im[(m_{e\mu}^{*})^{2}m_{\mu\tau}^{2}m_{ee}m_{\tau\tau}^{*}],
\end{equation}
where, $(k^{\dagger}k)_{11}=|a_{1}|^{2}+|c_{1}|^{2}$. From Eq. (\ref{eq57}), $\epsilon_{1}$ depends  on  physical phase   $i.e.$ arg$ [(m_{e\mu}^{*})^{2}m_{\mu\tau}^{2}m_{ee}m_{\tau\tau}^{*}]$.  
For type 2, one can obtain  $\epsilon_{1}$, simply by  the exchange of  $\mu\leftrightarrow\tau$ .\\
Similarly, with the help of Eq. \ref{eq56} and  $b_{1}, c_{1},c_{2},a_{2}$ in subsection 5.3,
 $\epsilon_{1}$ can be expressed  as
\begin{equation}\label{eq58}
\epsilon_{1}=-\frac{3}{16\pi v^{2}}\times \frac{M_{1}}{|m_{\mu\mu}||m_{ee}|^{2}(|m_{\mu \mu}|^{2}+|m_{\mu \tau}|^{2})} \times Im[(m_{\mu\tau}^{*})^{2}m_{e\tau}^{2}m_{\mu \mu}m_{ee}^{*}],
\end{equation} 
where, $(k^{\dagger}k)_{11}=|b_{1}|^{2}+|c_{1}|^{2}$. In case of type3, $\epsilon_{1}$ depends on  physical phase $i.e.$ arg $[(m_{\mu\tau}^{*})^{2}m_{e\tau}^{2}m_{\mu \mu}m_{ee}^{*}]$.
The result for type 4 can  simply be obtained through $\mu-\tau$ exchange symmetry.
From Eqs.(\ref{eq57}) and (\ref{eq58}), we conclude that CP-violating asymmetry $\epsilon_{1}$ requires the mismatch among the phases associated with  $m_{e\mu}, m_{\mu\tau}, m_{ee}$ and $ m_{\tau\tau}$  pertaining to $M_{\nu}$ for type1 ansatz, while, for type3, same holds true for  the  phases associated with  $m_{e\tau}, m_{\mu\tau}, m_{ee}$ and $ m_{\mu\mu}$.  This, in turn,  lead to net lepton number asymmetry, $Y_{L}\equiv \frac{nL}{s}=\frac{d\epsilon_{1}}{g^{*}}$, where $g^{*}=106.75$ corresponds to an effective number featuring the relativistic degree of freedom which contribute to the entropy $s$, and $d$ is  the dilution effects induced by the
lepton-number-violating wash-out processes \cite{leptogensis2}. The lepton number asymmetry $Y_{L}$ is finally converted into a net baryon number asymmetry $Y_{B}$ through the nonperturbative sphaleron processes \cite{YB}: $Y_{B} \equiv \frac{nB}{s} \approx −0.5Y_{L}$.

In addition to the phase dependence,   $\epsilon_{1} $ depends only on  $M_{1}$, for  $M_{2} >> M_{1}$.  Another  careful observation  reveal that $\epsilon_{1}$ for all the FGY ans\"{a}tze depends inversely on  $|m_{ee}|$. Therefore, the measurement of $|m_{ee}|$ through various neutrinoless double beta decay experiments is important for calculating the baryon asymmetry of Universe. In the following discussion, we shall see  how $\epsilon_{1}$  depends  explicitly on the  CP violating phases related to low energy.\\
With the help of Eqs.(\ref{eq51}) and (\ref{eq56})  we can arrive at following relations  
\begin{equation}\label{eq59}
\epsilon_{1}=-\frac{3}{16\pi v^{2}}M_{1}\Delta m_{12}^{2}\frac{\rm{Im} [c_{z}^{2}]}{\overline{m_{1}}},
\end{equation}
or 
 \begin{equation}\label{eq60}
\epsilon_{1}=+\frac{3}{16\pi v^{2}}M_{1}\Delta m_{12}^{2}\frac{\rm{Im} [s_{z}^{2}]}{\overline{m_{1}}},
\end{equation}
where, $\overline{m_{1}}= v(m_{1}|c_{z}|^{2}+m_{2}|s_{z}|^{2})$.\\
Using Eqs.(\ref{eq47},\ref{eq48}) and (\ref{eq59}, \ref{eq60}), $\epsilon_{1}$ is given as 
  \begin{equation}\label{eq61}
\epsilon_{1}=\mp \frac{3}{16\pi v^{2}} \frac{M_{1}\Delta m_{12}^{2}t_{12}t_{23}^{-1}c_{12}^{2}s_{13}}{\overline{m_{1}}}sin\delta,
\end{equation}
 for type 1 (minus) and type 4 (plus), respectively.\\
Similarly, using Eqs.(\ref{eq49}, \ref{eq50}) and (\ref{eq59}, \ref{eq60}), $\epsilon_{1}$ is  given as
  \begin{equation}\label{eq62}
\epsilon_{1}=\pm \frac{3}{16\pi v^{2}} \frac{M_{1}\Delta m_{12}^{2}t_{12}t_{23}c_{12}^{2}s_{13}}  {\overline{m_{1}}}sin\delta,
\end{equation}
for type 2(plus) and type 3(minus), respectively. These relations show the explicit dependence of lepton asymmetry on $\delta$. From Eqs. (\ref{eq44}) and (\ref{eq46}), it is clear that $sin\delta$ is directly proportional to $sin2\sigma$, implying that $\epsilon_{1}\propto sin2\sigma$. Therefore lepton asymmetry depends on the Majorana CP-violating phase $\sigma$. It is worthwhile to note that this phase parameter does not affect CP violation in neutrino oscillation, but it can be instrumental  in the scenarios of leptogenesis due to the lepton number violating and CP violating decays of the two heavy right handed Majorana neutrinos. The discussion also  remain consistent with  Ref. \cite{Hargya}.

\section{Summary and Conclusion}
In summary, we have considered the minimal seesaw model (MSM) augmented with two zero in the Dirac neutrino mass matrix.  Taking into account the four experimentally viable  ans\"{a}tze with inverted mass ordering,  we  construct the weak basis invariants (WB) relevant for leptogenesis in terms of low energy effective neutrino mass matrix elements, and then find the  necessary and sufficient conditions of CP conservation.  It is shown that  textures having three or more zeros lead to CP conservation. The CP violation at high energies for these ans\"{a}tze requires that phases among the low energy effective Majorana mass matrix elements are not fine tuned and, in addition,  the right handed Majorana neutrino masses $M_{1}$ and $M_{2}$ are non degenerate.    To extend our analysis further, we have explicitly shown the dependence of these CP odd invariants on Majorana CP violating phase ($\sigma$)  for each ans\"{a}tz, and find that $ \delta, \sigma=  \pm n\pi$, where n is a integer, holds for CP invariance in leptonic sector at high energy scale.\\  
 In the end we re-examine the implications of these interrelationships on leptogenesis. In this regard, we have shown the relations for CP violating asymmetry in terms of left handed Majorana neutrino mass matrix for all ans\"{a}tze. Further, it is shown that it's non-zero value depends on the mismatch among the phases associated with the elements of $M_{\nu}$. In addition,   for all ans\"{a}tze, CP  violating asymmetry  depends on effective neutrino mass, $|m_{ee}|$, related to neutrinoless double beta decay.
 
 In future long baseline experiments and neutrinoless double beta decay experiments, the precise determination of low energy parameters e.g.  CP violating phases($\delta$, $\sigma$), octant of $\theta_{23}$,  is critical to rule in or rule out the  FGY ans\"{a}tze.

\section*{Conflicts of Interest}
The author declares that there are no conflicts of interest regarding the publication of this paper.

\section*{Acknowledgment}
The author would like to thank the Principal of M. N. S. Government College,  Bhiwani, Haryana, India, for providing the necessary facilities to work. \\

\begin{table}
\begin{tiny}
\begin{center}
\begin{tabular}{|c|c|c|c|}
  \hline
 Cases of $M_{D}$ & $M_{\nu}$& CP-odd invariants & Phase Relationship for CP conservation  \\
  \hline
  &&&\\
  &&$I_{1}=M_{1}^{2}M_{2}^{2}(M_{2}^{2}-M_{1}^{2})Im\bigg[\frac{(m_{e\mu}^{*})^{2}m_{\mu\tau}^{2}}{m_{ee}^{*}m_{\tau\tau}}\bigg]$&\\
  &&$I_{2}=M_{1}M_{2}(M_{2}^{4}-M_{1}^{4})Im\bigg[\frac{(m_{e\mu}^{*})^{2}m_{\mu\tau}^{2}}{m_{ee}^{*}m_{\tau\tau}}\bigg]$&\\
   Type1& $
\begin{array}{ccc}
  \left(
\begin{array}{ccc}
  \frac{a_{1}^{2}}{M_{1}}& \frac{a_{1}b_{1}}{M_{1}}& 0 \\
  &\frac{b_{1}^{2}}{M_{1}}+\frac{b_{2}^{2}}{M_{2}} &\frac{b_{2}c_{2}}{M_{1}} \\
  && \frac{c_{2}^{2}}{M_{2}} \\
  \end{array}
  \right)
  \end{array}$ & $I_{3}=M_{1}^{3}M_{2}^{3}(M_{2}^{2}-M_{1}^{2})Im\bigg[\frac{(m_{e\mu}^{*})^{2}m_{\mu\tau}^{2}}{m_{ee}^{*}m_{\tau\tau}}\bigg]$& $arg(m_{ee})+2arg(m_{\mu\tau})=arg(m_{\tau\tau})+2arg(m_{e\mu})$\\  
  &&$I_{4}=m_{\mu}^{2}M_{1}^{2}M_{2}^{2}(M_{2}^{2}-M_{1}^{2})Im\bigg[\frac{(m_{e\mu}^{*})^{2}m_{\mu\tau}^{2}}{m_{ee}^{*}m_{\tau\tau}}\bigg]$&\\
&& $I_{5}=m_{\mu}^{2}M_{1}M_{2}(M_{2}^{4}-M_{1}^{4})Im\bigg[\frac{(m_{e\mu}^{*})^{2}m_{\mu\tau}^{2}}{m_{ee}^{*}m_{\tau\tau}}\bigg]$  &\\
&& $I_{6}=m_{\mu}^{2}M_{1}^{3}M_{2}^{3}(M_{2}^{2}-M_{1}^{2})Im\bigg[\frac{(m_{e\mu}^{*})^{2}m_{\mu\tau}^{2}}{m_{ee}^{*}m_{\tau\tau}}\bigg]$ &\\

   \hline
   &&&\\
   && $I_{1}=M_{1}^{2}M_{2}^{2}(M_{2}^{2}-M_{1}^{2})Im\bigg[\frac{(m_{e\tau}^{*})^{2}m_{\mu\tau}^{2}}{m_{ee}^{*}m_{\tau\tau}}\bigg]$&\\
   &&$I_{2}=M_{1}M_{2}(M_{2}^{4}-M_{1}^{4})Im\bigg[\frac{(m_{e\tau}^{*})^{2}m_{\mu\tau}^{2}}{m_{ee}^{*}m_{\tau\tau}}\bigg]$&\\
   Type2 &$
\left(
\begin{array}{ccc}
  \frac{a_{1}^{2}}{M_{1}}& 0& \frac{a_{1}c_{1}}{M_{1}} \\
  &\frac{b_{2}^{2}}{M_{2}} &\frac{b_{2}c_{2}}{M_{2}} \\
  &&\frac{c_{1}^{2}}{M_{1}}+\frac{c_{2}^{2}}{M_{2}}  \\
  \end{array}
  \right)
 $  & $I_{3}=M_{1}^{3}M_{2}^{3}(M_{2}^{2}-M_{1}^{2})Im\bigg[\frac{(m_{e\tau}^{*})^{2}m_{\mu\tau}^{2}}{m_{ee}^{*}m_{\tau\tau}}\bigg]$  & $arg(m_{ee})+2arg(m_{\mu\tau})=arg(m_{\tau\tau})+2arg(m_{e\mu})$ \\
  &&$I_{4}=m_{\tau}^{2}M_{1}^{2}M_{2}^{2}(M_{2}^{2}-M_{1}^{2})Im\bigg[\frac{(m_{e\tau}^{*})^{2}m_{\mu\tau}^{2}}{m_{ee}^{*}m_{\tau\tau}}\bigg]$&\\
&& $I_{5}=m_{\tau}^{2}M_{1}M_{2}(M_{2}^{4}-M_{1}^{4})Im\bigg[\frac{(m_{e\tau}^{*})^{2}m_{\mu\tau}^{2}}{m_{ee}^{*}m_{\tau\tau}}\bigg]$  &\\
&& $I_{6}=m_{\tau}^{2}M_{1}^{3}M_{2}^{3}(M_{2}^{2}-M_{1}^{2})Im\bigg[\frac{(m_{e\tau}^{*})^{2}m_{\mu\tau}^{2}}{m_{ee}^{*}m_{\tau\tau}}\bigg]$ &\\
   \hline
   &&&\\
   && $I_{1}=M_{1}^{2}M_{2}^{2}(M_{2}^{2}-M_{1}^{2})Im\bigg[\frac{(m_{\mu\tau}^{*})^{2}m_{e\tau}^{2}}{m_{\mu \mu}^{*}m_{ee}}\bigg]$&\\
   &&$I_{2}=M_{1}M_{2}(M_{2}^{4}-M_{1}^{4})Im\bigg[\frac{(m_{\mu\tau}^{*})^{2}m_{e\tau}^{2}}{m_{\mu \mu}^{*}m_{ee}}\bigg]$&\\
   Type3&$\left(
\begin{array}{ccc}
  \frac{a_{2}^{2}}{M_{2}}& 0& \frac{a_{2}c_{2}}{M_{2}} \\
  &\frac{b_{1}^{2}}{M_{1}} &\frac{b_{1}c_{1}}{M_{2}} \\
  &&\frac{c_{1}^{2}}{M_{1}}+\frac{a_{2}c_{2}}{M_{2}}  \\
  \end{array}
  \right)$  &  $I_{3}=M_{1}^{3}M_{2}^{3}(M_{2}^{2}-M_{1}^{2})Im\bigg[\frac{(m_{\mu\tau}^{*})^{2}m_{e\tau}^{2}}{m_{\mu \mu}^{*}m_{ee}}\bigg]$  & $arg(m_{\mu\mu})+2arg(m_{e\tau})=arg(m_{ee})+2arg(m_{\mu\tau})$ \\
  &&$I_{4}=m_{\tau}^{2}M_{1}^{2}M_{2}^{2}(M_{2}^{2}-M_{1}^{2})Im\bigg[\frac{(m_{\mu\tau}^{*})^{2}m_{e\tau}^{2}}{m_{\mu \mu}^{*}m_{ee}}\bigg]$&\\
&& $I_{5}=m_{\tau}^{2}M_{1}M_{2}(M_{2}^{4}-M_{1}^{4})Im\bigg[\frac{(m_{\mu\tau}^{*})^{2}m_{e\tau}^{2}}{m_{\mu \mu}^{*}m_{ee}}\bigg]$  &\\
&& $I_{6}=m_{\tau}^{2}M_{1}^{3}M_{2}^{3}(M_{2}^{2}-M_{1}^{2})Im\bigg[\frac{(m_{\mu\tau}^{*})^{2}m_{e\tau}^{2}}{m_{\mu \mu}^{*}m_{ee}}\bigg]$ &\\
  \hline
  &&&\\
   &&$I_{1}=M_{1}^{2}M_{2}^{2}(M_{2}^{2}-M_{1}^{2})Im\bigg[\frac{(m_{\mu\tau}^{*})^{2}m_{e\mu}^{2}}{m_{\tau \tau}^{*}m_{ee}}\bigg]$ & \\
   &&$I_{2}=M_{1}M_{2}(M_{2}^{4}-M_{1}^{4})Im\bigg[\frac{(m_{\mu\tau}^{*})^{2}m_{e\mu}^{2}}{m_{\tau \tau}^{*}m_{ee}}\bigg]$&\\
  Type4 &$\left(
\begin{array}{ccc}
  \frac{a_{2}^{2}}{M_{2}}& \frac{a_{2}b_{2}}{M_{2}}&0  \\
  &\frac{b_{1}^{2}}{M_{1}}+\frac{b_{2}^{2}}{M_{2}} &\frac{b_{1}c_{1}}{M_{2}} \\
  &&\frac{c_{1}^{2}}{M_{1}}  \\
  \end{array}
  \right)$ &  $I_{3}=M_{1}^{3}M_{2}^{3}(M_{2}^{2}-M_{1}^{2})Im\bigg[\frac{(m_{\mu\tau}^{*})^{2}m_{e\mu}^{2}}{m_{\tau \tau}^{*}m_{ee}}\bigg]$  & $arg(m_{\tau\tau})+2arg(m_{e\mu})=arg(m_{ee})+2arg(m_{\mu\tau})$ \\
  &&$I_{4}=m_{\mu}^{2}M_{1}^{2}M_{2}^{2}(M_{2}^{2}-M_{1}^{2})Im\bigg[\frac{(m_{\mu\tau}^{*})^{2}m_{e\mu}^{2}}{m_{\tau \tau}^{*}m_{ee}}\bigg]$&\\
&& $I_{5}=m_{\mu}^{2}M_{1}M_{2}(M_{2}^{4}-M_{1}^{4})Im\bigg[\frac{(m_{\mu\tau}^{*})^{2}m_{e\mu}^{2}}{m_{\tau \tau}^{*}m_{ee}}\bigg]$  &\\
&& $I_{6}=m_{\mu}^{2}M_{1}^{3}M_{2}^{3}(M_{2}^{2}-M_{1}^{2})Im\bigg[\frac{(m_{\mu\tau}^{*})^{2}m_{e\mu}^{2}}{m_{\tau \tau}^{*}m_{ee}}\bigg]$ &\\
  
\hline
\end{tabular}
\caption{\label{tab1} The structure of effective Majorana mass term ($M_{\nu}$), and the rephasing invariants $I_{1,2,3,4,5,6}$ as well as  the necessary and sufficient conditions for CP invariance is given corresponding to each FGY ans\"{a}tz.}
\end{center}
\end{tiny}
\end{table}

\end{document}